\documentclass[pre,groupedaddress,superscriptaddress,twocolumn,showpacs,reprint,notitlepage]{revtex4}

\usepackage{amsfonts}
\usepackage{amsmath}
\usepackage{amssymb}
\usepackage{graphicx}
\usepackage{hyperref}
\usepackage{color}
\usepackage{mathrsfs}
\usepackage{bbm}
\usepackage{subfigure}
\usepackage{times,txfonts}
\usepackage{isomath}
\usepackage{units}
\usepackage{nicefrac}

\DeclareMathAlphabet{\mathpzc}{OT1}{pzc}{m}{it}

\newcommand{\ignore}[1]{}
\newcommand{\nobibentry}[1]{{\let\nocite\ignore\bibentry{#1}}}
\newcommand{\bibfnamefont}[1]{#1}
\newcommand{\bibnamefont}[1]{#1}

\newcommand{\vect}[1]{\vectorsym{#1}} 
\newcommand{\ten}[1]{\tensorsym{#1}} 

\begin{document}

\title{Gaussian tripartite entanglement out of equilibrium}

\author{Antonio A. Valido}
\email{aavalido@ull.es}
\affiliation{IUdEA: Instituto Universitario de Estudios Avanzados, Universidad de La Laguna, La Laguna, 38203 Spain}
\affiliation{Dpto. F\'{\i}sica Fundamental II, Universidad de La Laguna, 38203 Spain}

\author{Luis A. Correa}
\email{lacorrea@ull.es}
\affiliation{IUdEA: Instituto Universitario de Estudios Avanzados, Universidad de La Laguna, La Laguna, 38203 Spain}
\affiliation{Dpto. F\'{\i}sica Fundamental, Experimental, Electr\'{o}nica y
Sistemas, Universidad de La Laguna, La Laguna, 38203 Spain}

\author{Daniel Alonso}
\email{dalonso@ull.es}
\affiliation{IUdEA: Instituto Universitario de Estudios Avanzados, Universidad de La Laguna, La Laguna, 38203 Spain}
\affiliation{Dpto. F\'{\i}sica Fundamental, Experimental, Electr\'{o}nica y
Sistemas, Universidad de La Laguna, La Laguna, 38203 Spain}

\pacs{03.65.Yz, 03.67.Mn, 03.67.Bg, 42.50.Lc}

\begin{abstract}
The stationary \textit{multipartite} entanglement between three interacting harmonic oscillators subjected to decoherence is analyzed in the largely unexplored non-equilibrium strong dissipation regime. We compute the exact asymptotic Gaussian state of the system and elucidate its separability properties, qualitatively assessing the regions of the space of parameters in which fully inseparable states are generated. Interestingly, the sharing structure of bipartite entanglement is seen to degrade as dissipation increases even for very low temperatures, at which the system approaches its ground state. We also find that establishing stationary energy currents across the harmonic chain does not correspond with the build-up of biseparable steady states, which relates instead just to the relative intensity of thermal fluctuations.
\end{abstract}

\date{\today}
\maketitle

\section{Introduction}
Entangled states of continuous variable (CV) systems have come to occupy a prominent position in quantum technologies \cite{braunstein20051} for both experimental and theoretical convenience. On the experimental side, the high degree of control in the preparation, manipulation and measurement of Gaussian CV states \cite{weedbrook20121} in a range of quantum physical supports including optical cavities, trapped ions \citep{leibfried20031} or nanomechanical devices \citep{ekinci20051}, makes them ideal for the efficient implementation of quantum information protocols. In particular, entangled CV \textit{multipartite} Gaussian states are a valuable resource for communication schemes involving many parties \cite{telenetwork,telecloning,adesso2006newjphys}, whose quantum-enhanced performance has been already demonstrated in experiments \cite{telenetworkexp,telecloningexp}. 

This \textit{outperformance} over classical protocols crucially relies on the amount and distribution of the entanglement shared by the multiple `modes', which makes the precise quantification of multipartite entanglement a matter of paramount importance. The general assessment of entanglement even in low-dimensional quantum systems remains an open and challenging problem to date \cite{horodecki2009a,bennett20111} and yet tremendous progress has been made towards its characterization in the CV Gaussian multipartite scenario \cite{giedke20011,adesso20071,adesso20121}. This fact, combined with the simple mathematical description that CV multi-mode Gaussian states enjoy, further highlights their practical convenience.

Unfortunately, entanglement is very fragile to the unavoidable decorrelating external environments and therefore, the successful implementation of quantum technologies with CVs should start with a complete understanding of noise and dissipation, so that they may be avoided or eventually engineered to protect quantum coherences. In this line, a number of recent works have extensively analyzed the dynamics and asymptotic properties of bimodal entanglement in CV Gaussian states under realistic models of noise and dissipation \cite{eisert20041,plenio20041,paz20081,paz2009pra,maniscalco2009common,maniscalco2009commonindependent,galve2010pra,galve2010prl,dechiara2011distant,ludwig20101,correa20121}. Concretely, the stationary two-mode entanglement under weak correlated and uncorrelated local noise was addressed in \cite{paz20081,maniscalco2009common,maniscalco2009commonindependent} for identical oscillators, and in \cite{paz2009pra,galve2010pra} for the non-resonant case. Moreover, the problem may be solved exactly once one abandons the assumption of weak interaction between system and environment, thus making it possible to probe into the strongly non-Markovian and non-equilibrium regimes \cite{ludwig20101,correa20121}. In contrast, much less is known about noise and dissipation in the CV Gaussian multipartite scenario \cite{paris2004bose,ferraro20052,adesso20061,xiang20091,gao_xiang20101,li20111,galve2013tripartite} where, to our knowledge, all available results are limited by either the weak dissipation or equilibration assumptions. 

The present paper aims to study multipartite stationary entanglement in the little-studied non-equilibrium strongly dissipative regime, through the extension of the exact techniques of \cite{correa20121}. We focus on the stationary Gaussian states that result from the contact of an interacting three-mode CV system with three local structured heat baths. A rich physical picture is gained by preparing the baths at generally different equilibrium temperatures, thus inducing steady-state energy transport. Endowed with all the versatility of an exact unconstrained stationary solution, we address the question whether robust tripartite entangled states may be generated out of equilibrium. As we shall see below, we can answer in the positive. 

More precisely, we take three (generally non-resonant) modes arranged in an open chain with linear nearest-neighbour interactions and locally dissipating into uncorrelated Ohmic baths. We are then able to compute the \emph{exact} Gaussian steady state of the system, under the sole assumption of initially uncorrelated system and environmental degrees of freedom \cite{weiss1999}. Our model is particularly suited for the theoretical description of tripartite CV systems in which thermal relaxation is the main source of decoherence, as it may occur, for instance, to trapped ions in a Paul trap \cite{brown20111} or clamped interacting nanomechanical oscillators \cite{buks20021,cleland20021}.

Taking the exact steady state as starting point, we issue a comprehensive study of the tripartite entanglement distribution according to the classification introduced in \cite{giedke20011}. When the three equilibrium temperatures of the reservoirs are set to the same value and identical oscillators are considered, we observe the expected competition between decoherence and inter-oscillator coupling in the build-up of stationary tripartite entanglement. Most interestingly, we find limiting dissipation rates above which the ground state of the interacting oscillators switches from the `weak dissipation' fully inseparable phase into a `strong dissipation' \emph{bound entangled} phase, passing through an intermediate \emph{two-mode biseparable} stage. As we shall see, these changes in the entanglement-sharing structure occur as a consequence of the non-negligible renormalization effects introduced by the system-bath interaction, in spite of the vanishing thermal fluctuations.

Imposing a temperature gradient across the chain proves detrimental to the formation of robust fully inseparable states unless the system is set up in an asymmetrical configuration. Nevertheless, the resulting separability structure does not seem to depend on the stationary energy currents induced across the system, but rather, with the relative intensity of thermal fluctuations on each of the modes. 

Finally, we discuss how the asymptotic tripartite entanglement may be enhanced with a suitable choice of parameters leading to well separated time scales for the thermal fluctuations and the free dynamics of the interacting modes.

This paper is organized as follows: We start by introducing the microscopic model for the system, the baths and their dissipative interaction in Sec.~\ref{SECsystem}. The reduced dynamics of the oscillators is tackled via the generalized quantum Langevin equation, introduced in Sec.~\ref{SSECqle}, and solved in the stationary regime in Secs.~\ref{SSECformal} and \ref{SSECohmic}. For a detailed derivation of the closed formula of the exact steady state, the interested reader is directed to Appendix~\ref{SECappendix}. We then briefly review the classification criteria for tripartite entanglement in CV Gaussian states in Sec.~\ref{SECentanglement}, and apply them to the steady states of our system in Sec.~\ref{SECresults}: The separability properties in the case of identical equilibrium temperatures are discussed in Sec.~\ref{SSECequilibrium}, and the results on the steady-state entanglement under a temperature gradient are presented in Sec.~\ref{SSECnonequilibrium}. Finally, in Sec.~\ref{SECconclusion}, we summarize and draw our conclusions.

\section{The system\label{SECsystem}}
\begin{figure*}[ht!]
\begin{center}
\includegraphics[width=0.9\textwidth]{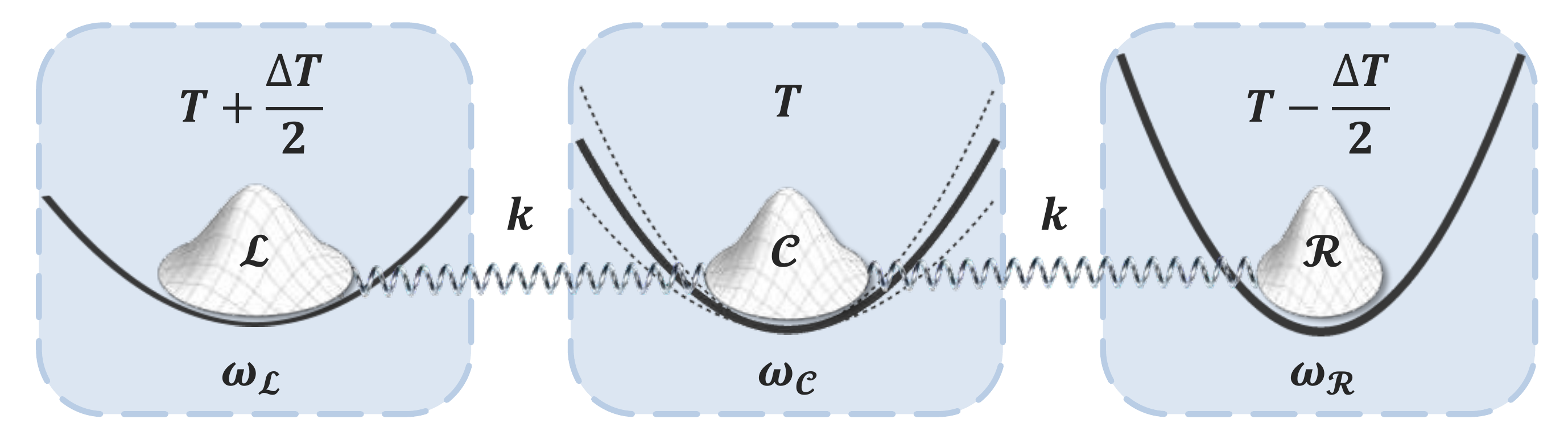}
\end{center}
\caption{(color online). Schematic representation of our tripartite CV system comprised of non-resonant modes mechanically coupled via nearest-neighbour linear interactions of strength $k$. Each oscillator dissipates at a rate $\gamma$ into its local bath, at temperatures $T_\mathcal{C}=T$ and $T_{\mathcal{L},\,\mathcal{R}}=T\pm\Delta T/2$, where $\Delta T\in [0,2T)$ so that a temperature gradient may be established across the system. 
\label{Fig1}}
\end{figure*}
As already mentioned, our system consist of three quantum harmonic oscillators, labelled by $\alpha\in\{\mathcal{L},\mathcal{C},\mathcal{R}\}$ after `left', `center' and `right', respectively. They have bare oscillation frequencies $\omega_\alpha$ and equal mass $m$ is assumed:
\begin{equation}
H_{S0}=\sum_\alpha  \frac{p_{\alpha}^{2}}{2m}+\frac{1}{2}m\omega_{\alpha}^{2}x_{\alpha}^{2}.
\label{HS0}\end{equation}
Here $x_\alpha$ and $p_\alpha$ stand for the corresponding position and momentum operators. We connect the oscillators through a generic \emph{quadratic} interaction term of the form
\begin{equation}
H_{SI}=\frac{1}{2}\sum_{\alpha\beta}x_\alpha V_{\alpha\beta}~x_\beta,
\label{HSI}\end{equation}
where $V_{\alpha\beta}$ are the entries of an Hermitian \textit{interaction matrix} $\ten{V}$. In particular, we shall arrange the oscillators in an open chain with nearest neighbour interactions of strength $k$ connecting $\mathcal{L}\leftrightarrow\mathcal{C}$ and $\mathcal{C}\leftrightarrow\mathcal{R}$, that is (see Fig.~\ref{Fig1} below)
\begin{equation}
\ten V=\left(\begin{array}{ccc}
 k & -k & 0 \\
-k & 2k & -k\\
 0 & -k & k
\end{array}\right).
\label{Vab}\end{equation}
We address the local dissipation mechanism with the paradigmatic Caldeira-Legget model \cite{caldeira1983,weiss1999}. Therefore, three independent bosonic reservoirs are introduced, also labeled $\alpha\in\{\mathcal{L},\mathcal{C},\mathcal{R}\}$, comprised of non-interacting modes $\{\mathpzc{q}_{\alpha\mu},\mathpzc{p}_{\alpha\mu}\}$ linearly coupled to their local oscillator $\{x_\alpha,p_\alpha\}$ with strength $g_{\alpha\mu}$:
\begin{equation}
H_{SB}=\sum_{\alpha\mu}  \frac{\mathpzc{p}_{\alpha\mu}^{2}}{2m_{\alpha\mu}}+\frac{1}{2}m_{\alpha\mu}\omega_{\alpha\mu}^{2}\left(\mathpzc{q}_{\alpha\mu}-\frac{g_{\alpha\mu}}{m_{\alpha\mu}\omega_{\alpha\mu}^{2}}x_{\alpha}\right)
^{2}.
\label{HSB}\end{equation}
Apart from the free Hamiltonian of the reservoirs and their linear interaction with the system (i.e. the terms of the form $g_{\alpha\mu} x_{\alpha}\mathpzc{q}_{\alpha\mu}$), Eq.~\eqref{HSB} also explicitly includes the \emph{renormalization term}
\begin{equation}
H_{R}=\sum_{\alpha\mu}\frac{g_{\alpha\mu}^2}{2m_{\alpha\mu}\omega_{\alpha\mu}^{2}}x_{\alpha}^2,
\label{renormalization}\end{equation}
which is necessary in order to compensate the distortion exerted by the system-bath coupling on $H_{S0}$ \cite{weiss1999}. The effects of this term only start to become relevant as the system-baths interaction grows stronger \cite{correa20121}. The coupling constants $g_{\alpha\mu}$ define the spectral densities
\begin{equation}
J_\alpha(\omega)\equiv\pi\sum_\mu \frac{g_{\alpha\mu}^2}{2m_{\alpha\mu}\omega_{\alpha\mu}}~\delta(\omega-\omega_{\alpha\mu}),
\label{spectraldens}\end{equation}
which receive a phenomenological functional form suitable for a correct description of dissipation. In particular, in Sec.~\ref{SSECohmic}, we shall consider Ohmic spectral densities with Lorentz-Drude high frequency cutoff
\begin{equation}
J_\alpha(\omega)=\frac{m\gamma_\alpha\omega}{1+\omega^2/\omega_c^2},
\label{ohmic}\end{equation}
where $\gamma_\alpha$ stands for the \textit{dissipation rate}, and carries the order of magnitude of the system-bath interaction, and $\omega_c$ is the cutoff frequency, that places a lower bound in the characteristic time scale of the thermal fluctuations of the baths \cite{breuer}. 

We initialize system and environment as $\varrho_0=\rho_{0}\otimes\big(\bigotimes_\alpha\tau_\alpha\big)$, where $\rho_{0}$ is any state of the three oscillators, $\tau_\alpha=\mathcal{Z}_\alpha^{-1}e^{-H_{B_\alpha}/k_B T_\alpha}$ is a (Gaussian) thermal equilibrium state of reservoir $\alpha$ at temperature $T_\alpha$, and where $k_B$ denotes the Boltzmann constant. The normalization factors are $\mathcal{Z}_\alpha\equiv\text{tr}~\{e^{-H_{B_\alpha}/k_B T_\alpha}\}$ and $H_{B_\alpha}$ stands for the free Hamiltonian of the corresponding reservoir. The linearity of the system's effective dynamics, guaranteed by the overall linear Hamiltonian and the `Gaussianity' of the baths, leads to Gaussian reduced stationary states $\rho_\infty=\text{tr}_B~\{\varrho_\infty\}$ \cite{garbert1984oscillator}.

Any Gaussian three-mode state is fully determined (up to local displacements) by its second order moments, arranged in the $6\times 6$ covariance matrix
\begin{equation}
\boldsymbol{\sigma}\equiv\left(\begin{array}{c | c} \ten C_{\vect X \vect X}(0) & \ten C_{\vect X \vect P}(0) \\ \hline \ten C_{\vect P \vect X}(0) & \ten C_{\vect P \vect P}(0) \end{array}\right).
\label{covariance}\end{equation}
The $3\times 3$ blocks $\ten{C}_{\vect A \vect B}(0)$ are defined as
\begin{equation}
\ten{C}_{\vect A \vect B}(t-t')\equiv\frac{1}{2}\langle \vect A(t) \vect B^T(t')+\vect B(t') \vect A^T(t)\rangle_{\rho_0},
\label{correlation}\end{equation}
where $\vect A,\vect B\in\{\vect X,\vect P\}$ and $\vect X=\{x_\mathcal{L},x_\mathcal{C},x_\mathcal{R}\}^T$, $\vect P=\{p_\mathcal{L},p_\mathcal{C},p_\mathcal{R}\}^T$ are column vectors collecting position and momentum operators of the modes. 

\section{Exact stationary states\label{SECstationary}}

\subsection{Generalized quantum Langevin equation\label{SSECqle}}

We shall now calculate the stationary matrices $\ten C_{\vect A \vect B}(0)$ and thus, the steady state of the system, by making use of the generalized quantum Langevin equation (QLE) formalism \cite{weiss1999}, which is widespreadly used in the study of quantum Brownian motion \cite{hanggi20051}. The QLE follows from the elimination of the environment in the Heisenberg equations of motion for $x_\alpha(t)$ and $p_\alpha(t)$, and may be compactly written as
\begin{equation}
\ten{M} \ddot{\vect X}+ \ten \phi \vect X=\vect{\eta}(t)+\frac{1}{\hbar}{\displaystyle\int\limits_{-\infty}^{t}}d\tau~{\ten{\chi}}(t-\tau)\vect{X}(\tau).
\label{langevin}\end{equation}
Note that this equation does not rely on any approximations and therefore, it remains valid in all regimes of parameters. We remark as well that we took the initial condition $\varrho_0$ at $t_0\rightarrow-\infty$ so that for any finite $t$, it already describes the asymptotic properties of the system. 

The $3\times 3$ matrix $\ten M$ is diagonal and carries the masses of the oscillators $M_{\alpha\beta}=m\delta_{\alpha\beta}$, where $\delta_{\alpha\beta}$ stands for Kronecker delta. The effective potential is encoded in $\phi_{\alpha\beta}=m\omega_\alpha^2\delta_{\alpha\beta}+V_{\alpha\beta}+2m~\Delta\Omega_{\alpha}\delta_{\alpha\beta}$, where the frequency shift
\begin{equation}
m~\Delta\Omega_{\alpha}\equiv\frac{1}{\pi}\int_0^\infty d\omega ~\frac{J_\alpha(\omega)}{\omega},
\label{freqshift}\end{equation}
directly follows from the renormalization term of Eq.~\eqref{renormalization}. 

In addition to the free dynamics of the interacting oscillators, Eq.~\eqref{langevin} also accounts for decoherence: On the one hand, the oscillators are locally driven by the stochastic quantum forces $\eta_\alpha(t)$ that enclose the effects of the thermal noise. These form the column vector $\vect \eta(t)$. On the other hand, the last term on the right-hand side stands for a `friction memory kernel' or `generalized susceptibility' and describes dissipation. Since the three baths are uncorrelated, the $3\times 3$ susceptibility matrix $\ten \chi$ has elements
\begin{equation}
\chi_{\alpha\beta}(t)\equiv\delta_{\alpha\beta}~\Theta(t)~\frac{2\hbar}{\pi}\int_0^\infty d\omega~J_\alpha(\omega)\sin{\omega t},
\label{memorykernel}\end{equation}
where $\Theta(t)$ stands for the Heaviside step function. Thermal noise and friction are connected via the Kubo relation
\begin{equation}
\ten \chi (t-t')=-i	\Theta(t-t')\left\langle \vect\eta(t)\vect\eta^{T}(t')-\vect\eta(t')\vect\eta^{T}(t) \right\rangle_B,
\label{fluctdiss}\end{equation}
where $\left\langle A \right\rangle_B\equiv\text{tr}~\{A \bigotimes_\alpha\tau_\alpha\}$ denotes an average over the environmental degrees of freedom.

\subsection{Formal stationary solution\label{SSECformal}}

Quite generically, the matrices $\ten C_{\vect A \vect B}(t)$ may be extracted from Eq.~\eqref{langevin} by taking its Fourier transform $\tilde{f}(\omega)\equiv\int dt~e^{i\omega t} f(t)$. One thus arrives to the linear expression
\begin{equation}\label{xw}
\tilde{\vect X}(\omega) = \ten \alpha \left( \omega \right)  \tilde{\vect \eta}\left( \omega \right),
\end{equation}
where the complex matrix $\alpha(\omega)$ is defined as
\begin{equation}
\ten \alpha(\omega)\equiv-\left(\omega^{2} \ten M - \ten \phi + \frac{1}{\hbar}~\tilde{\ten \chi}\left(\omega \right)\right)^{-1},
\label{alpha}\end{equation}
and the Fourier transform $\tilde{\ten\chi}(\omega)$ of the generalized susceptibility matrix has elements such that
\label{imgensuscept}\begin{equation}
-\frac{\text{Im}~\tilde{\chi}_{\alpha\alpha}(\omega)}{\hbar}=J_\alpha(\omega)\,\Theta(\omega)-J_\alpha(-\omega)\,\Theta(-\omega).
\end{equation}
The causality argument that renders $\chi_{\alpha\alpha}(t)=0$ $\forall~t<0$ also ensures that $\tilde{\chi}_{\alpha\alpha}(\omega)$ is analytic in the upper-half plane of complex frequencies \cite{weiss1999}. By virtue of the Kramers-Kronig relations we then have
\begin{equation}
\text{Re}~\tilde{\chi}_{\alpha\alpha}(\omega)=\mathcal{P}\int_{-\infty}^{\infty}\frac{d\omega'}{\pi}~\frac{\text{Im}~\tilde{\chi}_{\alpha\alpha}(\omega')}{\omega'-\omega},
\label{realgensuscept}\end{equation}
where $\mathcal{P}$ stands for the principal value of the integral. Let us now introduce the notation
\begin{equation}
\Gamma_{\alpha}(\omega)\equiv -\frac{\text{Im}~\tilde{\chi}_{\alpha\alpha}(\omega)}{\hbar}\coth{\frac{\hbar\omega}{2k_B T_\alpha}},
\label{powerspectrum}\end{equation}
for the \textit{symmetrized power spectrum} of the quantum stochastic force $\eta_\alpha(t)$ \cite{ludwig20101,correa20121}, and the vector $\ten \Gamma(\omega)\equiv\{\Gamma_\mathcal{L}(\omega),\Gamma_\mathcal{C}(\omega),\Gamma_\mathcal{R}(\omega)\}^T$. Then, the matrix $\ten C_{\vect X \vect X}(t)$ writes as
\begin{equation}
\ten C_{\vect X \vect X}(t)= \hbar \int \frac{d\omega}{2 \pi} e^{-i \omega t} \ten \alpha(\omega) \ten \Gamma(\omega)\ten\alpha(-\omega)^{T},
\label{Cxx}\end{equation}
while the remaining correlations are: 
\begin{equation}
\ten C_{\vect P \vect P}(t) = \hbar\,m^2 {\displaystyle\int} \frac{d\omega}{2 \pi} ~\omega^2\,e^{-i \omega t}\ten \alpha(\omega) \ten \Gamma(\omega)\ten\alpha(-\omega)^{T} \label{Cpp},
\end{equation}
and $\ten C_{\vect X \vect P}(t)=\ten C_{\vect P \vect X}(t)$
\begin{equation}
\ten C_{\vect X \vect P}(t) = i \hbar\,m {\displaystyle\int} \frac{d\omega}{2 \pi} ~\omega\,e^{-i \omega t}\ten \alpha(\omega) \ten \Gamma(\omega)\ten\alpha(-\omega)^{T} \label{Cxp}.
\end{equation}
Eqs.~\eqref{alpha}-\eqref{Cxp} thus formally provide the desired exact stationary states of the system for arbitrary spectral densities $J_\alpha(\omega)$.

\subsection{Stationary solution for Ohmic baths\label{SSECohmic}}

As already anticipated, in order to compute the steady state from Eqs.~\eqref{alpha}-\eqref{Cxp}, we will restrict ourselves to the Ohmic spectral densities of Eq.~\eqref{ohmic} and further assume symmetric dissipation rates $\gamma_\alpha=\gamma$. In this case, $ \tilde{\ten\chi}(\omega)$ reduces to
\begin{equation}
\tilde{\chi}_{\alpha\beta}(\omega)=\delta_{\alpha\beta}~\frac{m\hbar\gamma\omega_c^2}{i\omega-\omega_c},
\label{susceptibilityohmic}\end{equation}
which gives $\alpha(\omega)$ and $\Gamma_\alpha(\omega)$ by immediate substitution into Eqs.~\eqref{alpha} and \eqref{powerspectrum}. Note that the frequency shift of Eq.~\eqref{freqshift} is now $\Delta\Omega_\alpha=\gamma\omega_c/2$. 

It is indeed possible to carry out the integration in Eqs.~\eqref{Cxx}-\eqref{Cxp} and get closed formulas for the exact correlations by means of contour integration in the plane of complex frequencies, as in \cite{riseborough1895exact}. Unfortunately, little can be gained from the cumbersome expressions that result, neither from the physical, nor from the practical point of view. Their discussion is hence postponed until Appendix \ref{SECappendix}, and in what follows, we shall evaluate of Eqs.~\eqref{Cxx}-\eqref{Cxp} numerically.

In the next section, we briefly review the basic tools to be employed in the characterization of the entanglement distribution in the stationary states of our system.

\section{Gaussian tripartite entanglement\label{SECentanglement}}

As already mentioned, the precise quantification of genuine multipartite entanglement in general mixed states, still proves challenging \cite{bennett20111,horodecki2009a} even in the simplest case of tripartite systems. For instance, when dealing with qubits, quantities that prove to be \textit{bona fide} measures in the bipartite scenario, such as the concurrence \cite{bennet1996eof} or the negativity \cite{vidal2002a}, have to be replaced with a suitable entanglement mononotone that additionally satisfies the Coffman-Kundu-Wootters (CKW) \textit{monogamy} inequality, like the residual \textit{tangle}, computed from the convex roof of the squared concurrence \cite{coffman2000CKW}.

In complete analogy, a continuous variable residual tangle, or (Gaussian) \textit{cotangle}, was introduced in \cite{adesso20061} that satisfies the CKW inequality for all three-mode Gaussian states. It follows from the infimum of the squared logarithmic negativity \cite{vidal2002a} taken over all possible (Gaussian) pure-state decompositions of $\rho$. Alternatively, a monogamous Gaussian entanglement measure may also be defined in terms of the R\'{e}nyi-2 entropy \cite{adesso20121}.

As a bipartite entanglement measure, the (logarithmic) negativity exploits the positivity-of-the-partial-transpose (PPT) separability criterion \cite{peres1996ppt,horodecki1996ppt} which turns out to be not only necessary, but also \emph{sufficient} for all $1\times n$ multi-mode Gaussian states \cite{werner20001gaussian1n}. Therefore, even if the (logarithmic) negativity fails to faithfully account for genuine multipartite correlations, the PPT criterion does allow for a qualitative description of the distribution of Gaussian entanglement in a three-mode CV system, according to the number of non-separable bipartitions out of the three possible. We shall denote them as $\mathcal{L}\vert(\mathcal{C}\mathcal{R})$, $\mathcal{C}\vert(\mathcal{L}\mathcal{R})$ and $\mathcal{R}\vert(\mathcal{L}\mathcal{C})$. This entails the following classification for tripartite Gaussian states, as introduced in \cite{giedke20011}:

\begin{enumerate}
\item[\textit{C1}.] \textit{Fully inseparable states}, that are not separable in any of the bipartitions.
\item[\textit{C2}.] \textit{One-mode biseparable states}, which are separable only in one out of the three possible bipartitions. 
\item[\textit{C3}.] \textit{Two-mode biseparable states}, for which now two of the bipartitions are separable.
\item[\textit{C4}.] \textit{Three-mode biseparable} or bound entangled states, which are separable under all bipartitions, but cannot be written as a mixture of product states only.
\item[\textit{C5}.] \textit{Fully separable states}, that unlike those of \textit{C4}, \emph{can} be written as a mixture of product states.
\end{enumerate}

In order to distinguish between the PPT-equivalent classes \textit{C4} and \textit{C5}, we make use of the criterion for full separability of \cite{giedke20011}. In what follows, rather than attempting to quantify genuine tripartite entanglement, we resort to the previous qualitative characterization and apply it to the exact stationary states of our system.

\section{Results and discussion\label{SECresults}}

Finally, we are in a position to analyze the distribution of the stationary tripartite entanglement classes in the space of parameters of the system. Even if Eqs.~\eqref{alpha}-\eqref{Cxp} are not underpinned by any restrictive assumptions, we shall focus on the low temperature regime, which is optimal for the build up of entanglement, and exploit our steady-state solution to probe into the strongly dissipative regime. 

We shall also restrict to low effective inter-oscillator coupling strengths $k$, as \textit{strong} couplings are rather unrealistic in experiments. This translates into $k/m\Omega^2\ll 1$, where $\Omega\sim\omega_\alpha$. Indeed, by noting that $\tau_k\sim m\Omega/k$ is a characteristic time for energy transport across the system when isolated from the environment, it becomes clear that the condition $k/m\Omega^2\ll 1$ amounts to a separation of time scales $\tau_k\gg\Omega^{-1}$ that renders transport inefficient. Consequently, the typical time scale governing the closed evolution of the whole interacting system may be approximated as $\tau_S\sim\Omega^{-1}$.

In the study of quantum Brownian motion, one usually assumes \emph{fast} thermal fluctuations ($\tau_B\sim \hbar/k_B T\ll\tau_S$, $\tau_B\ll\tau_D$) as compared with the free evolution and the dissipation time $\tau_D\sim\gamma^{-1}$ \cite{breuer}. On the contrary, we shall work with relatively \emph{low} temperatures and \emph{strong} dissipation rates ($k_B T/\hbar \lesssim \Omega$, $k_B T/\hbar\sim\gamma$) so that the system is much more insensitive to noise. In this regime, picking a cutoff frequency $\omega_c$ of the order of $\Omega$ gives rise to \textit{non-perturbative} renormalization frequency shifts $\Delta\Omega=\gamma\omega_c/2$ that should be expected to become relevant. It is also important to note that under strong dissipation, the stationary states of the system are generally not of \textit{thermal equilibrium} (Gibbs states) \cite{garbert1984oscillator,riseborough1895exact,haake19851}, even when the temperatures of the local baths coincide and no steady-state energy transport is established.

Under these conditions, the stationary tripartite entanglement is studied in absence of energy currents through the system (Sec.~\ref{SSECequilibrium}), and when the equilibrium temperatures of the baths are arranged in a gradient (Sec.~\ref{SSECnonequilibrium}).

\begin{figure}[t]
\includegraphics[width=0.9\columnwidth]{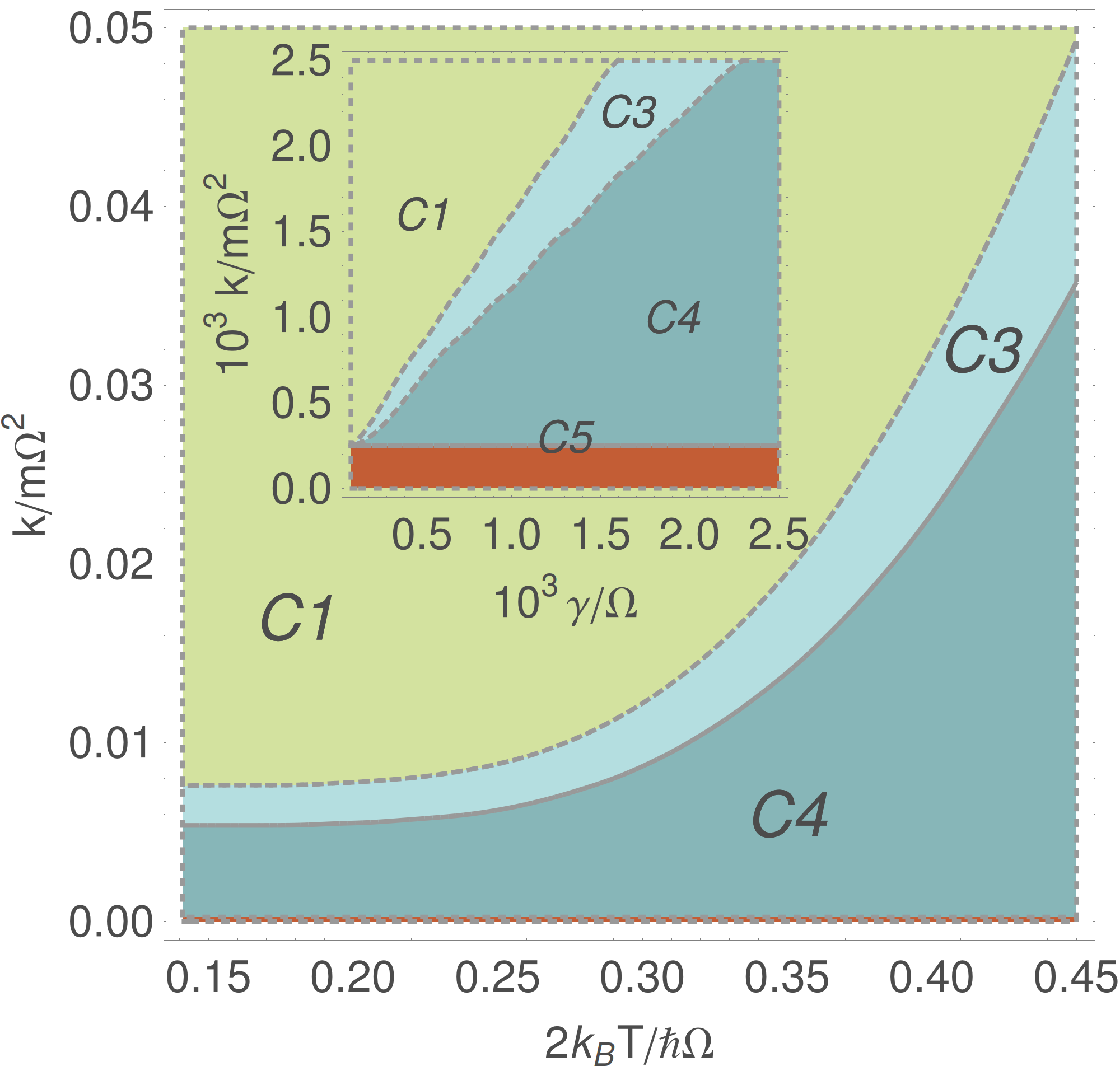} 
\caption{(color online). Phase diagram with Gaussian tripartite entanglement classes as a function of the inter-oscillator coupling strength $k$ and the temperature of the baths $T_\alpha=T$ for $\omega_\alpha=\Omega$. The dissipation rate was fixed to $\gamma=10^{-2}\Omega$, while the cutoff frequency is $\omega_c=50\Omega$. For sufficiently weak coupling, the stationary state lies within the fully separable class (\textit{C5}), which is almost imperceptible at the bottom of the plot. In the inset, the tripartite entanglement classes are shown as a function of the interaction strength $k$ and the dissipation rate $\gamma$, at a very low temperature of just $2k_BT/\hbar\Omega=0.05$. We observe that for any $k$ above a temperature-dependent threshold, the ground state undergoes a transition from the fully inseparable phase, characteristic of low dissipation, to a bound entangled phase (\textit{C4}), passing through an intermediate two-mode biseparable stage (\textit{C3}) as the dissipation rate is increased.\label{Fig2}}
\end{figure}

\begin{figure*}[ht!]
\begin{center}
\subfigure{\includegraphics[width=0.33\textwidth]{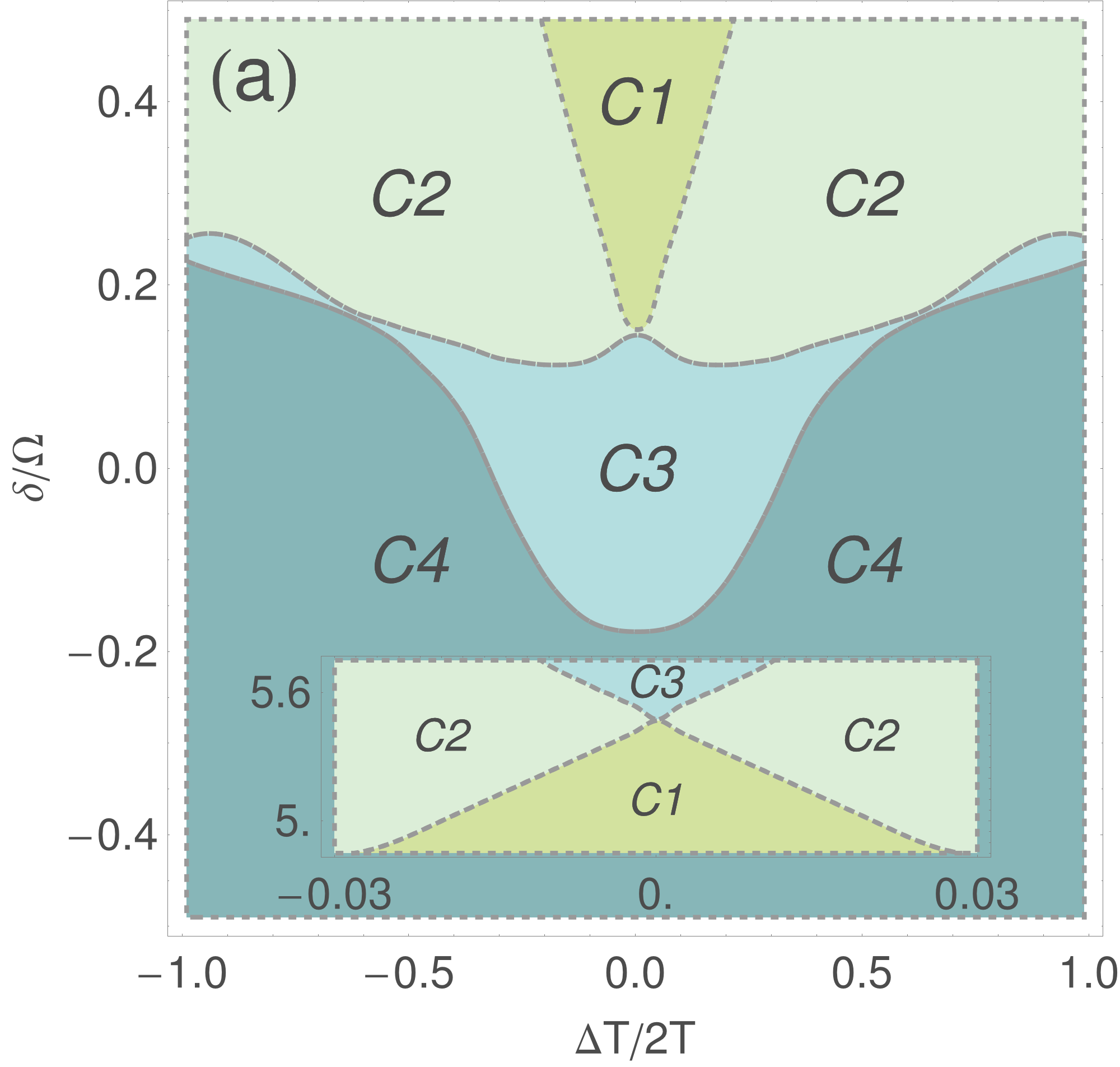}\label{Fig3a}}
\subfigure{\includegraphics[width=0.33\textwidth]{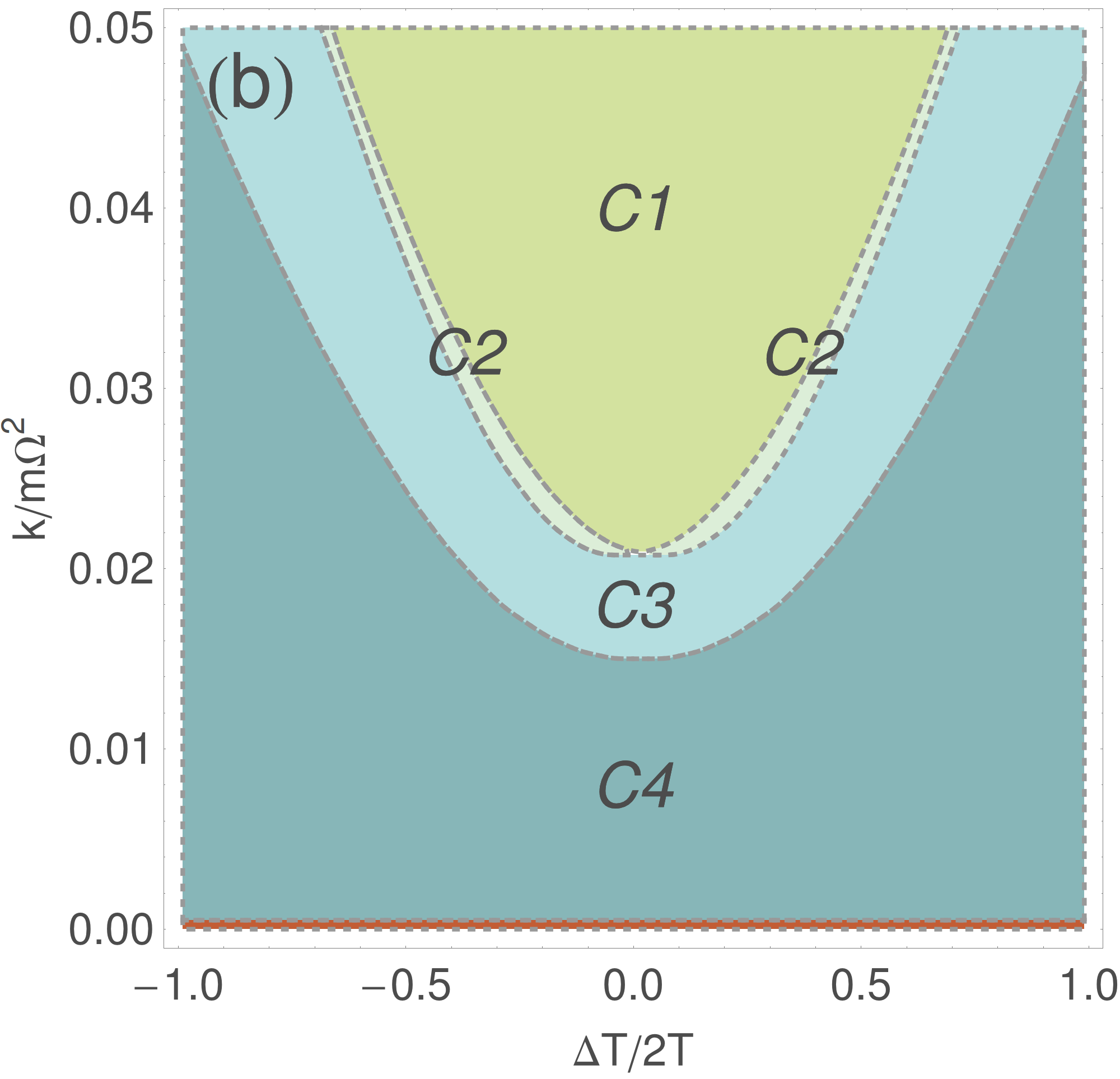}\label{Fig3b}} 
\subfigure{\includegraphics[width=0.33\textwidth]{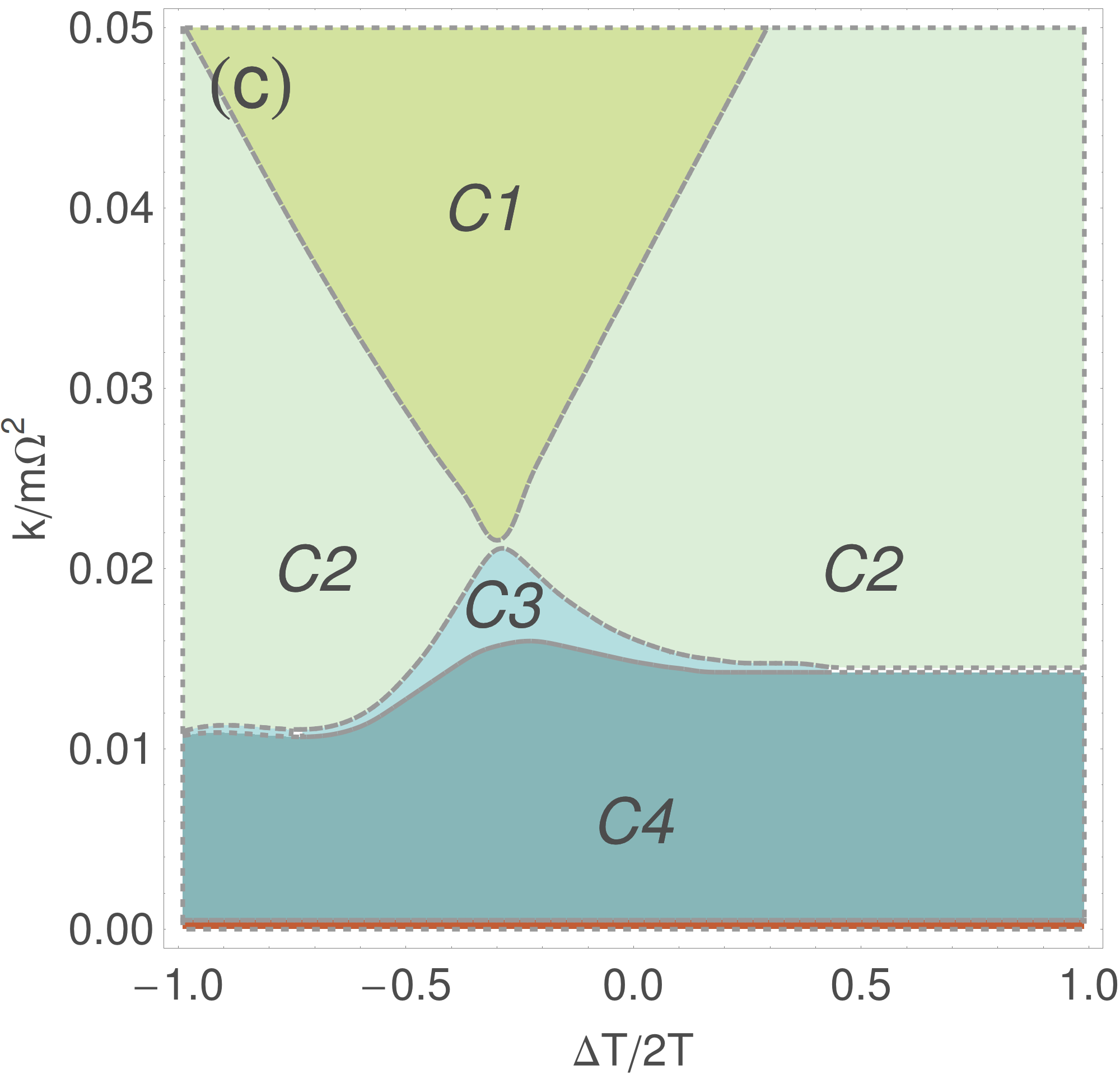}\label{Fig3c}} 
\caption{(color online). Stationary Gaussian tripartite entanglement classes versus: (a) $\delta$ and $\Delta T$ for coupling strength $k/m\Omega^2=0.05$ and $\omega_{\mathcal{L},\,\mathcal{R}}=\Omega$, $\omega_\mathcal{C}=\Omega+\delta$, (b) $k$ and $\Delta T$ for $\omega_\alpha=\Omega$, and (c) $k$ and $\Delta T$ for $\omega_\mathcal{L}=\Omega$, $\omega_\mathcal{C}=2\Omega$ and $\omega_\mathcal{R}=3\Omega$ (see discussion in Sec.~\ref{SSECnonequilibrium}). All three figures share the same average temperature $2k_BT/\hbar\Omega=0.35$ and the same dissipation rate $\gamma$ and cutoff $\omega_c$ as in Fig.~\ref{Fig2}. In the inset of Fig.~\ref{Fig3a}, we zoom in around $\Delta T=0$ for large detuning $\delta$, and observe how the fully inseparable phase (\textit{C1}) is a \emph{connected} region in the $\delta$--$\Delta T$ space.\label{Fig3}}
\end{center}
\end{figure*}

\subsection{Identical equilibrium temperatures\label{SSECequilibrium}}

We shall start by taking resonant frequencies $\omega_\alpha=\Omega$ and $\Delta T=0$ (see caption of Fig.~\ref{Fig1}). The tripartite entanglement class of the resulting stationary states is plotted in Fig.~\ref{Fig2} as a function of the coupling strength $k$ and the equilibrium temperatures $T$ of the baths. Not surprisingly, the higher the temperatures, the higher the corresponding coupling $k$ that is required to keep the system in a fully inseparable state (\textit{C1}). Note as well that one mode biseparable states (\textit{C2}) do not build up asymptotically in this configuration.

In fact, the stationary entanglement in the bipartition $\mathcal{C}\vert(\mathcal{L}\mathcal{R})$ proves more resilient to noise than in either $\mathcal{L}\vert(\mathcal{C}\mathcal{R})$ or $\mathcal{R}\vert(\mathcal{L}\mathcal{C})$. This is obviously due to our choice of potential $\ten V$ in Eq.~\eqref{Vab}, that only puts mode $\mathcal{C}$ in direct interaction with the remaining two. Now, given that in this configuration the system is invariant under the exchange $\mathcal{L}\leftrightarrow\mathcal{R}$, its stationary states must be bisymmetric and, therefore, as the temperatures increase, steady-state entanglement in bipartitions $\mathcal{L}\vert(\mathcal{C}\mathcal{R})$ and $\mathcal{R}\vert(\mathcal{L}\mathcal{C})$ must disappear \emph{jointly}, which entails a direct transition from \textit{C1} to \textit{C3}. Increasing the temperatures further, the system also becomes separable with respect to $\mathcal{C}\vert(\mathcal{L}\mathcal{R})$, thus giving rise to stationary bound entangled states (\textit{C4}). Even though class \textit{C5} only appears for extremely low coupling in Fig.~\ref{Fig2}, at any given $k$ there exist a temperature $T$ above which the steady states become fully separable \cite{anders20081}.

Most interestingly, in the inset of Fig.~\ref{Fig2} we can see how the separability properties of the ground state (GS) of the chain depend on $k$ and $\gamma$: For any $k$ above a temperature-dependent threshold $k^{T}_{\min}$ (in the figure $k^{T}_{\min}\simeq 2.5\times 10^{-3} k/m\Omega^2$), there exist dissipation rates at which the GS undergoes transitions \textit{C1}$\rightarrow$\textit{C3} and \textit{C3}$\rightarrow$\textit{C4}. On the contrary, for $k<k_{\min}^{T}$, it remains in the fully inseparable phase \textit{C5} regardless of the dissipation strength. The sharing structure of bipartite entanglement in the GS of a harmonic chain thus depends on $\gamma$ when \textit{decohering} far from the Born-Markov regime.

This can be, at least, qualitatively understood by recalling that the system Hamiltonian $H_{S0}+H_{R}$ \emph{includes} the renormalization term of Eq.~\eqref{HSB}, that amounts to a shift on the frequencies $\Omega^2\mapsto\Omega^2_r\equiv\Omega^2+2~\Delta \Omega$. Hence, one may argue that the effective coupling strength $k/m\Omega_r^2$ decreases as the dissipation rate grows, thus potentially downgrading the GS to an entanglement class of higher separability.

\subsection{Temperature gradient across the system\label{SSECnonequilibrium}}

We now arrange the baths in a temperature gradient by allowing for $\Delta T\neq 0$ (see Fig.~\ref{Fig1}) so that stationary energy transport may be established across the harmonic chain. Let us first consider $2k_BT/\hbar\Omega=0.35$, $\omega_{\mathcal{L},\,\mathcal{R}}=\Omega$ and $\omega_\mathcal{C}=\Omega+\delta$. This configuration is invariant with respect to the combined exchange of $\mathcal{L}\leftrightarrow\mathcal{R}$ and $\Delta T\leftrightarrow-\Delta T$ and thus, the distribution of entanglement phases must be symmetric about $\Delta T = 0$, as seen in Figs.~\ref{Fig3a} and \ref{Fig3b}. 

In Fig.~\ref{Fig3a} we fix $k/m\Omega^2=0.05$ and plot the entanglement classes as a function of $\delta$ and $\Delta T$. First, notice that one-mode biseparable stationary states (\textit{C2}) do build up, now that the symmetry argument invoked in Sec.~\ref{SSECequilibrium} is not applicable. 

One sees as well that in general, whenever $\omega_\mathcal{C}$ increases, the free dynamics of the central mode becomes more insensitive to noise since $k_B T/\hbar\omega_\mathcal{C}$ decreases. This helps to reduce the stationary \textit{biseparability} and eventually yields fully inseparable states (\textit{C1}). However, as illustrated in the inset, very large values of $\omega_\mathcal{C}$ may also cause an effective decoupling of the central mode from the rest as $k/m\omega_\mathcal{C}^2$ becomes smaller. In other words, \emph{given} a fixed interaction $k$, fully inseparable stationary states may be generated by tuning the frequencies to a compromise between shielding the system from thermal noise and keeping the effective interaction between its modes \textit{sufficiently strong}. 

Finally, note that arranging the baths in a temperature gradient proves detrimental to the asymptotic formation of states in any of the bipartite entangled classes (\textit{C1}--\textit{C3}). This seems to occur due to the intensification of thermal noise at the hot end of the chain rather than as a consequence of the stationary energy currents established across the system. We illustrate this point further in Figs.~\ref{Fig3b} and \ref{Fig3c}, where $k$ and $\Delta T$ are taken as the free parameters.

In Fig.~\ref{Fig3b} we consider resonant modes ($\delta=0$), while in Fig.~\ref{Fig3c} the oscillators are set up in the asymmetrical configuration: $\omega_\mathcal{L}=\Omega$, $\omega_\mathcal{C}=2\Omega$ and $\omega_\mathcal{R}=3\Omega$. In the first case, keeping the steady state within the fully inseparable class requires stronger couplings as the temperature gradient increases in either direction. On the contrary, the asymmetric setting of Fig.~\ref{Fig3c} favors the formation of class \textit{C1} steady states at \textit{moderate} negative temperature gradients, as these provide the low frequency mode $\mathcal{L}$ with the lowest temperature ($T-\left\vert\Delta T\right\vert$) and the high frequency mode $\mathcal{R}$ with the highest one ($T+\left\vert\Delta T\right\vert$), which optimally shields the system from thermal noise. 

It is also noticeable how the one-mode biseparable class (\textit{C2}) takes over bound entangled steady states (\textit{C4}) in Fig.~\ref{Fig3c} as contrasted with Fig.~\ref{Fig3b}, even though it may be seen that the magnitude of the stationary energy currents \cite{dhar2006transport} are comparable in either case. This observation further suggests that the build-up of steady-state quantum correlations might indeed not share a causal relation with the efficient transport of energy at microscopic scale, as already pointed out in different contexts such as excitation transfer in biological systems \cite{plenio2008networks}, thermal conduction in spin chains \cite{wu20111} or the optimized performance of quantum refrigerators \cite{correa2013performance}.

\section{Conclusions\label{SECconclusion}}

We have addressed the qualitative classification of the bipartite entanglement distribution across three linearly coupled harmonic oscillators dissipating into independent structured baths. By making use of the quantum Langevin equation formalism, we were able to compute their exact stationary Gaussian states and then, issue a comprehensive analysis of the different entanglement classes that build-up asymptotically in terms of the parameters of the system. It is important to remark that this approach is not limited by the customary assumptions of equilibrium and/or weak-memoryless system-bath interactions, so that it allows to probe into the largely unexplored non-equilibrium strong dissipation regime.

Interestingly, we saw how the ground state of the harmonic chain undergoes structural transitions between different schemes of entanglement sharing, increasing its bipartite separability as the dissipation grows stronger. This is a direct consequence of the non-negligible back action of the system-bath coupling on the system itself.

It was also noted that inducing stationary energy transport by means of a temperature gradient is generally detrimental to the formation of fully inseparable steady states due to the more intense thermal fluctuations at the hot end of the system. The resulting stationary energy currents do not seem to correlate to the asymptotic formation of biseparable states. 

We finally discussed how a suitable choice of frequencies may shield the system from thermal noise while keeping the effective inter-oscillator coupling strong enough, so that potentially useful fully inseparable states may build up asymptotically in spite of the strong decoherence.

As it was already pointed out, our model is appropriate for the theoretical description of a range of systems of interest in quantum technologies, especially arrays of interacting nanomechanical resonators. Indeed, considering typical frequencies $\Omega$ in the range of $1$ MHz and masses $m$ around $10^{-15}$ kg, the region of the space of parameters probed in our numerics may be achieved in present-day experiments.

One could also think of applying the powerful exact techniques illustrated here to the study of steady-state multipartite entanglement under the action of \emph{correlated} thermal noise in a more realistic structured bath of spatial dimension greater than one. This problem is worthy of detailed study and will be considered elsewhere.

\section*{Acknowledgments}
The authors warmly thank A. Ruiz for reading and extensively commenting on the manuscript, and J. P. Palao, G. Adesso, D. Girolami, G. De Chiara, S. Kohler and N. Garc\'{i}a Marco for fruitful discussions and helpful criticism. L.A.C. wants to thank N. Ramos Garc\'{i}a \textit{in memoriam} for his unconditional support through the years. This project was funded by the Spanish MICINN (Grant No. FIS2010-19998) and the European Union (FEDER). A.A.V.  and L.A.C. acknowledge the Canary Islands Government for financial support through the ACIISI fellowships (85\% cofunded by European Social Fund).

\appendix

\section{Analytical expression for the covariance matrix\label{SECappendix}}

As already mentioned in Sec.~\ref{SSECohmic}, in order to get an analytical expressions for e.g., Eq.~\eqref{Cxx}, one can use the customary toolbox of complex analysis to explicitly carry out the integration. Therefore, complete knowledge about the roots $z_i$ of the denominator of the integrand is required. Let us start by alternatively writing $\ten C_{\vect X \vect X}(0)$ as
\begin{multline}
\frac{\left[\ten C_{\vect X \vect X} (0)\right]_{\alpha\delta}}{m\hbar\gamma\omega_c^2} \\= \sum_\beta \int \frac{d\omega}{2 \pi} \frac{\text{adj}\left[\ten F(\omega)\right]_{\alpha\beta}~\text{adj}\left[\ten F(\omega)^*\right]_{\beta\delta}}{\vert \ten F(\omega) \vert~\vert \ten F(\omega)^* \vert}~\omega\coth{\frac{\hbar\omega}{2k_B T_\beta}},
\label{derCxx1}\end{multline}
where the matrix $\ten F(\omega)$ is defined as $\left[\ten F(\omega)\right]_{\alpha\beta}\equiv(\omega_c-i\omega)\,[\ten\alpha^{-1}(\omega)]_{\alpha\beta}$. The notation $\text{adj}\left[\ten F(\omega)\right]=\vert\ten F(\omega)\vert~\ten F(\omega)^{-1}$ stands for the \textit{adjugate} matrix of $\ten F(\omega)$, and the asterisk represents conjugate transposition. Note that from Eq.~\eqref{susceptibilityohmic} it follows that $\ten\alpha(-\omega)^T=\ten\alpha(\omega)^*$.

The denominator of Eq.~\eqref{derCxx1} is a real polynomial of degree eighteen comprised of the determinants $\vert\ten F(\omega) \vert$ and $\vert\ten F(\omega)^* \vert$, which are complex polynomials of degree nine. Provided that $\ten F(\omega)$  is diagonalizable, $\vert\ten F(\omega) \vert$ may be written as the product of three polynomials of degree three, and therefore, its roots can be analytically worked out, even if the resulting expressions are rather involved. When it comes to the multiplicity of those complex roots, it can be checked that they are all simple for our choice of interaction potential in Eq.~\eqref{Vab}. We shall label them so that $\{z_1,\cdots,z_9\}$ lie in the lower half plane of complex frequencies (and $\{z_{10},\cdots,z_{18}\}=\{\overline{z}_1,\cdots,\overline{z}_9\}$ are their corresponding complex conjugates).

We may now decompose the integrand of Eq.~\eqref{derCxx1} into partial fractions as
\begin{widetext}
\begin{multline}
\frac{\text{adj}\left[\ten F(\omega)\right]_{\alpha\beta}~\text{adj}\left[\ten F(\omega)^*\right]_{\beta\delta}~\omega\coth{\frac{\hbar\omega}{2k_B T_\beta}}}{\vert \ten F(\omega) \vert~\vert \ten F(\omega)^* \vert} 
= \frac{1}{m^ 6}\sum_{j=1}^9 \frac{ \text{adj}\left[\ten F(\omega)\right]_{\alpha\beta}\text{adj}\left[\ten F(\omega)^*\right]_{\beta\delta}~\omega\coth{\frac{\hbar\omega}{2k_B T_\beta}}}{2i~\text{Im}~z_j \prod_{k \neq j}(z_j-z_k)\prod_{k \neq j}(z_j-\overline{z}_k)}~\frac{1}{\omega-z_j}\\
-\frac{1}{m^ 6}\sum_{j=1}^9\frac{ \text{adj}\left[\ten F(\omega)\right]_{\alpha\beta}\text{adj}\left[\ten F(\omega)^*\right]_{\beta\delta}~\omega\coth{\frac{\hbar\omega}{2k_B T_\beta}}}{2i~\text{Im}~z_j \prod_{k \neq j}(\overline{z}_j-z_k)\prod_{k \neq j}(\overline{z}_j-\overline{z}_k)}~\frac{1}{\omega-\overline{z}_j},
\label{partialfractions}\end{multline}
\end{widetext}
with $k\in\{1,\cdots,9\}$. We shall also make use of the identity
\begin{equation}
\coth{x}=\frac{1}{x}+\frac{1}{i\pi}\left[\psi\left(1+\frac{i x}{\pi} \right)-\psi\left(1-\frac{i x}{\pi} \right)\right],
\label{cothdigamma}\end{equation}
where $\psi(z)$ stands for the \textit{digamma} or \textit{psi-fuction}, i.e. the logarithmic derivative of Euler's gamma function $\psi(z)\equiv d\ln{\Gamma(z)/dz}$ \cite{abrahmowitz1972handbook}. 

Combining Eq.~\eqref{cothdigamma} with \eqref{partialfractions}, Eq.~\eqref{derCxx1} may be evaluated by making the analytical continuation of the integrand into the plane of complex frequencies and calculating residues. Notice that the extended function $\psi\left(1\pm i z/\pi \right)$ has simple poles along the entire positive (negative) imaginary axis. We shall choose integration contours either in the lower or upper plane for each of the resulting terms in Eq.~\eqref{derCxx1}, such that those non-analyticities are avoided. The elements of the correlation $\ten C_{\vect X \vect X}(0)$ thus result in
\begin{widetext}
\begin{equation}
\left[\ten C_{\vect X \vect X} (0)\right]_{\alpha\delta} = \frac{\hbar\gamma\omega_c^2}{m^5}\sum_{\beta}\sum_{j=1}^9
\left[\frac{k_B T_{\beta}}{2\hbar~\text{Im}~z_j}-\frac{2\text{Re}~z_j}{\pi~\text{Im}~z_j}~\text{Im}~\psi\left(1 + i \frac{\hbar z_j}{2\pi k_B T_\beta}\right)\right]~\text{Re}~\frac{ \text{adj}\left[\ten F(z_j)\right]_{\alpha\beta}\text{adj}\left[\ten F(z_j)^*\right]_{\beta\delta}}{\prod_{k \neq j}(z_j-z_k)\prod_{k \neq j}(z_j-\overline{z}_k)}.
\end{equation}
\end{widetext}
Similarly, $\ten C_{\vect P \vect P}(0)$ may be computed from Eq.~\eqref{Cpp} to yield
\begin{widetext}
\begin{equation}
\left[\ten C_{\vect P \vect P} (0)\right]_{\alpha\delta} = \frac{\hbar\gamma\omega_c^ 2}{m^3} \sum_{\beta}\sum_{j=1}^9
\left[\frac{k_B T_{\beta}~\text{Re}~z_j^2}{2\hbar~\text{Im}~z_j}-\frac{2\text{Re}~z_j^3}{\pi~\text{Im}~z_j}~\text{Im}~\psi\left(1 + i \frac{\hbar z_j}{2\pi k_B T_\beta}\right)\right]~\text{Re}~\frac{ \text{adj}\left[\ten F(z_j)\right]_{\alpha\beta}\text{adj}\left[\ten F(z_j)^*\right]_{\beta\delta}}{\prod_{k \neq j}(z_j-z_k)\prod_{k \neq j}(z_j-\overline{z}_k)},
\end{equation}
\end{widetext}
and finally, Eq.~\eqref{Cxp} translates into
\begin{widetext}
\begin{equation}
\left[\ten C_{\vect X \vect P} (0)\right]_{\alpha\delta} =- \frac{\hbar\gamma\omega_c^ 2}{m^4} \sum_{\beta}\sum_{j=1}^9
\left[\frac{k_B T_{\beta}~\text{Re}~z_j}{2\hbar~\text{Im}~z_j}-\frac{2\text{Re}~z_j^2}{\pi~\text{Im}~z_j}~\text{Im}~\psi\left(1 + i \frac{\hbar z_j}{2\pi k_B T_\beta}\right)\right]~\text{Im}~\frac{ \text{adj}\left[\ten F(z_j)\right]_{\alpha\beta}\text{adj}\left[\ten F(z_j)^*\right]_{\beta\delta}}{\prod_{k \neq j}(z_j-z_k)\prod_{k \neq j}(z_j-\overline{z}_k)},
\end{equation}
\end{widetext}
which provides us with the desired explicit formulas for the exact stationary Gaussian state of the system.

%


\begin{thebibliography}{43}%
\makeatletter
\providecommand \@ifxundefined [1]{%
 \@ifx{#1\undefined}
}%
\providecommand \@ifnum [1]{%
 \ifnum #1\expandafter \@firstoftwo
 \else \expandafter \@secondoftwo
 \fi
}%
\providecommand \@ifx [1]{%
 \ifx #1\expandafter \@firstoftwo
 \else \expandafter \@secondoftwo
 \fi
}%
\providecommand \natexlab [1]{#1}%
\providecommand \enquote  [1]{``#1''}%
\providecommand \bibnamefont  [1]{#1}%
\providecommand \bibfnamefont [1]{#1}%
\providecommand \citenamefont [1]{#1}%
\providecommand \href@noop [0]{\@secondoftwo}%
\providecommand \href [0]{\begingroup \@sanitize@url \@href}%
\providecommand \@href[1]{\@@startlink{#1}\@@href}%
\providecommand \@@href[1]{\endgroup#1\@@endlink}%
\providecommand \@sanitize@url [0]{\catcode `\\12\catcode `\$12\catcode
  `\&12\catcode `\#12\catcode `\^12\catcode `\_12\catcode `\%12\relax}%
\providecommand \@@startlink[1]{}%
\providecommand \@@endlink[0]{}%
\providecommand \url  [0]{\begingroup\@sanitize@url \@url }%
\providecommand \@url [1]{\endgroup\@href {#1}{\urlprefix }}%
\providecommand \urlprefix  [0]{URL }%
\providecommand \Eprint [0]{\href }%
\providecommand \doibase [0]{http://dx.doi.org/}%
\providecommand \selectlanguage [0]{\@gobble}%
\providecommand \bibinfo  [0]{\@secondoftwo}%
\providecommand \bibfield  [0]{\@secondoftwo}%
\providecommand \translation [1]{[#1]}%
\providecommand \BibitemOpen [0]{}%
\providecommand \bibitemStop [0]{}%
\providecommand \bibitemNoStop [0]{.\EOS\space}%
\providecommand \EOS [0]{\spacefactor3000\relax}%
\providecommand \BibitemShut  [1]{\csname bibitem#1\endcsname}%
\let\auto@bib@innerbib\@empty

\bibitem [{\citenamefont {Braunstein}\ and\ \citenamefont {van
  Loock}(2005)}]{braunstein20051}%
  \BibitemOpen
  \bibfield  {author} {\bibinfo {author} {\bibfnamefont {S.~L.}\ \bibnamefont
  {Braunstein}}\ and\ \bibinfo {author} {\bibfnamefont {P.}~\bibnamefont {van
  Loock}},\ } {\bibfield  {journal}
  {\bibinfo  {journal} {Rev. Mod. Phys.}\ }\textbf {\bibinfo {volume} {77}},\
  \bibinfo {pages} {513} (\bibinfo {year} {2005})}\BibitemShut {NoStop}%
\bibitem [{\citenamefont {Weedbrook}\ \emph {et~al.}(2012)\citenamefont
  {Weedbrook}, \citenamefont {Pirandola}, \citenamefont {Garc\'ia Patr\'on},
  \citenamefont {Cerf}, \citenamefont {Ralph}, \citenamefont {Shapiro},\ and\
  \citenamefont {Lloyd}}]{weedbrook20121}%
  \BibitemOpen
  \bibfield  {author} {\bibinfo {author} {\bibfnamefont {C.}~\bibnamefont
  {Weedbrook}}, \bibinfo {author} {\bibfnamefont {S.}~\bibnamefont
  {Pirandola}}, \bibinfo {author} {\bibfnamefont {R.}~\bibnamefont {Garc\'ia
  Patr\'on}}, \bibinfo {author} {\bibfnamefont {N.~J.}\ \bibnamefont {Cerf}},
  \bibinfo {author} {\bibfnamefont {T.~C.}\ \bibnamefont {Ralph}}, \bibinfo
  {author} {\bibfnamefont {J.~H.}\ \bibnamefont {Shapiro}}, \ and\ \bibinfo
  {author} {\bibfnamefont {S.}~\bibnamefont {Lloyd}},\ } {\bibfield  {journal} {\bibinfo  {journal} {Rev.
  Mod. Phys.}\ }\textbf {\bibinfo {volume} {84}},\ \bibinfo {pages} {621}
  (\bibinfo {year} {2012})}\BibitemShut {NoStop}%
\bibitem [{\citenamefont {Leibfried}\ \emph {et~al.}(2003)\citenamefont
  {Leibfried}, \citenamefont {Blatt}, \citenamefont {Monroe},\ and\
  \citenamefont {Wineland}}]{leibfried20031}%
  \BibitemOpen
  \bibfield  {author} {\bibinfo {author} {\bibfnamefont {D.}~\bibnamefont
  {Leibfried}}, \bibinfo {author} {\bibfnamefont {R.}~\bibnamefont {Blatt}},
  \bibinfo {author} {\bibfnamefont {C.}~\bibnamefont {Monroe}}, \ and\ \bibinfo
  {author} {\bibfnamefont {D.}~\bibnamefont {Wineland}},\ }\href@noop {}
  {\bibfield  {journal} {\bibinfo  {journal} {Rev. Mod. Phys.}\
  }\textbf {\bibinfo {volume} {75}},\ \bibinfo {pages} {281} (\bibinfo {year}
  {2003})}\BibitemShut {NoStop}%
\bibitem [{\citenamefont {Ekinci}\ and\ \citenamefont
  {Roukes}(2005)}]{ekinci20051}%
  \BibitemOpen
  \bibfield  {author} {\bibinfo {author} {\bibfnamefont {K.}~\bibnamefont
  {Ekinci}}\ and\ \bibinfo {author} {\bibfnamefont {M.}~\bibnamefont
  {Roukes}},\ }\href@noop {} {\bibfield  {journal} {\bibinfo  {journal} {Rev. Sci. Instrum.}\ }\textbf {\bibinfo {volume} {76}},\ \bibinfo
  {pages} {061101} (\bibinfo {year} {2005})}\BibitemShut {NoStop}%
\bibitem{telenetwork}
P. van Loock and S. L. Braunstein,
Phys. Rev. Lett. {\bf 84}, 3482 (2000).

\bibitem{telecloning}
P. van Loock and S. L. Braunstein,
Phys. Rev. Lett. {\bf 87}, 247901 (2001).

\bibitem{adesso2006newjphys}
G. Adesso, A. Serafini, and Fabrizio Illuminati,
New J. Phys. {\bf 9}, 60 (2006).

\bibitem{telenetworkexp}
H. Yonezawa, T. Aoki, and A. Furusawa,
Nature \textbf{431}, 430 (2004).

\bibitem{telecloningexp}
S. Koike, H. Takahashi, H. Yonezawa, N. Takei, S. L. Braunstein, T. Aoki, and A. Furusawa,
Phys. Rev. Lett. {\bf 96}, 060504 (2006).

\bibitem [{\citenamefont {Ferraro}\ \emph {et~al.}(2005)\citenamefont
  {Ferraro}, \citenamefont {Olivares},\ and\ \citenamefont
  {Paris}}]{ferraro20051}%
  \BibitemOpen
  \bibfield  {author} {\bibinfo {author} {\bibfnamefont {A.}~\bibnamefont
  {Ferraro}}, \bibinfo {author} {\bibfnamefont {S.}~\bibnamefont {Olivares}}, \
  and\ \bibinfo {author} {\bibfnamefont {M.}~\bibnamefont {Paris}},\
  }\href@noop {} {\emph {\bibinfo {title} {Gaussian states in quantum
  information}}}\ (\bibinfo  {publisher} {Bibliopolis},\ \bibinfo {year}
  {2005})\BibitemShut {NoStop}%
\bibitem [{\citenamefont {Horodecki}\ \emph {et~al.}(2009)\citenamefont
  {Horodecki}, \citenamefont {Horodecki}, \citenamefont {Horodecki},\ and\
  \citenamefont {Horodecki}}]{horodecki2009a}%
  \BibitemOpen
  \bibfield  {author} {\bibinfo {author} {\bibfnamefont {R.}~\bibnamefont
  {Horodecki}}, \bibinfo {author} {\bibfnamefont {P.}~\bibnamefont
  {Horodecki}}, \bibinfo {author} {\bibfnamefont {M.}~\bibnamefont
  {Horodecki}}, \ and\ \bibinfo {author} {\bibfnamefont {K.}~\bibnamefont
  {Horodecki}},\ } {\bibfield
  {journal} {\bibinfo  {journal} {Rev. Mod. Phys.}\ }\textbf {\bibinfo {volume}
  {81}},\ \bibinfo {pages} {865} (\bibinfo {year} {2009})}\BibitemShut
  {NoStop}%
\bibitem [{\citenamefont {Bennett}\ \emph {et~al.}(2011)\citenamefont
  {Bennett}, \citenamefont {Grudka}, \citenamefont {Horodecki}, \citenamefont
  {Horodecki},\ and\ \citenamefont {Horodecki}}]{bennett20111}%
  \BibitemOpen
  \bibfield  {author} {\bibinfo {author} {\bibfnamefont {C.~H.}\ \bibnamefont
  {Bennett}}, \bibinfo {author} {\bibfnamefont {A.}~\bibnamefont {Grudka}},
  \bibinfo {author} {\bibfnamefont {M.}~\bibnamefont {Horodecki}}, \bibinfo
  {author} {\bibfnamefont {P.}~\bibnamefont {Horodecki}}, \ and\ \bibinfo
  {author} {\bibfnamefont {R.}~\bibnamefont {Horodecki}},\ } {\bibfield  {journal} {\bibinfo  {journal} {Phys.
  Rev. A}\ }\textbf {\bibinfo {volume} {83}},\ \bibinfo {pages} {012312}
  (\bibinfo {year} {2011})}\BibitemShut {NoStop}%
\bibitem [{\citenamefont {Giedke}\ \emph {et~al.}(2001)\citenamefont {Giedke},
  \citenamefont {Kraus}, \citenamefont {Lewenstein},\ and\ \citenamefont
  {Cirac}}]{giedke20011}%
  \BibitemOpen
  \bibfield  {author} {\bibinfo {author} {\bibfnamefont {G.}~\bibnamefont
  {Giedke}}, \bibinfo {author} {\bibfnamefont {B.}~\bibnamefont {Kraus}},
  \bibinfo {author} {\bibfnamefont {M.}~\bibnamefont {Lewenstein}}, \ and\
  \bibinfo {author} {\bibfnamefont {J.~I.}\ \bibnamefont {Cirac}},\ } {\bibfield  {journal} {\bibinfo
  {journal} {Phys. Rev. A}\ }\textbf {\bibinfo {volume} {64}},\ \bibinfo
  {pages} {052303} (\bibinfo {year} {2001})}\BibitemShut {NoStop}%
\bibitem [{\citenamefont {Adesso}\ and\ \citenamefont
  {Illuminati}(2007)}]{adesso20071}%
  \BibitemOpen
  \bibfield  {author} {\bibinfo {author} {\bibfnamefont {G.}~\bibnamefont
  {Adesso}}\ and\ \bibinfo {author} {\bibfnamefont {F.}~\bibnamefont
  {Illuminati}},\ }\href@noop {} {\bibfield  {journal} {\bibinfo  {journal}
  {J. Phys. A}\ }\textbf {\bibinfo
  {volume} {40}},\ \bibinfo {pages} {7821} (\bibinfo {year}
  {2007})}\BibitemShut {NoStop}%
\bibitem [{\citenamefont {Adesso}\ \emph {et~al.}(2012)\citenamefont {Adesso},
  \citenamefont {Girolami},\ and\ \citenamefont {Serafini}}]{adesso20121}%
  \BibitemOpen
  \bibfield  {author} {\bibinfo {author} {\bibfnamefont {G.}~\bibnamefont
  {Adesso}}, \bibinfo {author} {\bibfnamefont {D.}~\bibnamefont {Girolami}}, \
  and\ \bibinfo {author} {\bibfnamefont {A.}~\bibnamefont {Serafini}},\ } {\bibfield  {journal} {\bibinfo
  {journal} {Phys. Rev. Lett.}\ }\textbf {\bibinfo {volume} {109}},\ \bibinfo
  {pages} {190502} (\bibinfo {year} {2012})}\BibitemShut {NoStop}%
\bibitem [{\citenamefont {Eisert}\ \emph {et~al.}(2004)\citenamefont {Eisert},
  \citenamefont {Plenio}, \citenamefont {Bose},\ and\ \citenamefont
  {Hartley}}]{eisert20041}%
  \BibitemOpen
  \bibfield  {author} {\bibinfo {author} {\bibfnamefont {J.}~\bibnamefont
  {Eisert}}, \bibinfo {author} {\bibfnamefont {M.}~\bibnamefont {Plenio}},
  \bibinfo {author} {\bibfnamefont {S.}~\bibnamefont {Bose}}, \ and\ \bibinfo
  {author} {\bibfnamefont {J.}~\bibnamefont {Hartley}},\ }\href@noop {}
  {\bibfield  {journal} {\bibinfo  {journal} {Phys. Rev. Lett.}\
  }\textbf {\bibinfo {volume} {93}},\ \bibinfo {pages} {190402} (\bibinfo
  {year} {2004})}\BibitemShut {NoStop}%
\bibitem [{\citenamefont {Plenio}\ \emph {et~al.}(2004)\citenamefont {Plenio},
  \citenamefont {Hartley},\ and\ \citenamefont {Eisert}}]{plenio20041}%
  \BibitemOpen
  \bibfield  {author} {\bibinfo {author} {\bibfnamefont {M.}~\bibnamefont
  {Plenio}}, \bibinfo {author} {\bibfnamefont {J.}~\bibnamefont {Hartley}}, \
  and\ \bibinfo {author} {\bibfnamefont {J.}~\bibnamefont {Eisert}},\
  }\href@noop {} {\bibfield  {journal} {\bibinfo  {journal} {New J. Phys.}\ }\textbf {\bibinfo {volume} {6}},\ \bibinfo {pages} {36} (\bibinfo
  {year} {2004})}\BibitemShut {NoStop}%
\bibitem [{\citenamefont {Paz}\ and\ \citenamefont
  {Roncaglia}(2008)}]{paz20081}%
  \BibitemOpen
  \bibfield  {author} {\bibinfo {author} {\bibfnamefont {J.~P.}\ \bibnamefont
  {Paz}}\ and\ \bibinfo {author} {\bibfnamefont {A.~J.}\ \bibnamefont
  {Roncaglia}},\ } {\bibfield
  {journal} {\bibinfo  {journal} {Phys. Rev. Lett.}\ }\textbf {\bibinfo
  {volume} {100}},\ \bibinfo {pages} {220401} (\bibinfo {year}
  {2008})}\BibitemShut {NoStop}%
\bibitem{paz2009pra}
J. P. Paz and A. J. Roncaglia, 
Phys. Rev. A {\bf 79}, 032102 (2009).

\bibitem{maniscalco2009common}
R. Vasile, S. Olivares, M. G. A. Paris, and S. Maniscalco,
Phys. Rev. A {\bf 80}, 062324 (2009).

\bibitem{maniscalco2009commonindependent}
R. Vasile, P. Giorda, S. Olivares, M. G. A. Paris, and S. Maniscalco,
Phys. Rev. A {\bf 82}, 012313 (2010).

\bibitem{galve2010pra}
F. Galve, G. L. Giorgi, R. Zambrini,
Phys. Rev. A {\bf 81}, 062117 (2010).

\bibitem{galve2010prl}
F. Galve, L. A. Pach\'{o}n, and D. Zueco,
Phys. Rev. Lett. {\bf 105}, 180501 (2010).  

\bibitem{dechiara2011distant}
A. Wolf, G. De Chiara, E. Kajari, E. Lutz, and G. Morigi, 
Europhys. Lett. \textbf{95}, 60008 (2011). 
\bibitem [{\citenamefont {Ludwig}\ \emph {et~al.}(2010)\citenamefont {Ludwig},
  \citenamefont {Hammerer},\ and\ \citenamefont {Marquardt}}]{ludwig20101}%
  \BibitemOpen
  \bibfield  {author} {\bibinfo {author} {\bibfnamefont {M.}~\bibnamefont
  {Ludwig}}, \bibinfo {author} {\bibfnamefont {K.}~\bibnamefont {Hammerer}}, \
  and\ \bibinfo {author} {\bibfnamefont {F.}~\bibnamefont {Marquardt}},\ } {\bibfield  {journal} {\bibinfo
  {journal} {Phys. Rev. A}\ }\textbf {\bibinfo {volume} {82}},\ \bibinfo
  {pages} {012333} (\bibinfo {year} {2010})}\BibitemShut {NoStop}%
\bibitem [{\citenamefont {Correa}\ \emph {et~al.}(2012)\citenamefont {Correa},
  \citenamefont {Valido},\ and\ \citenamefont {Alonso}}]{correa20121}%
  \BibitemOpen
  \bibfield  {author} {\bibinfo {author} {\bibfnamefont {L.~A.}\ \bibnamefont
  {Correa}}, \bibinfo {author} {\bibfnamefont {A.~A.}\ \bibnamefont {Valido}},
  \ and\ \bibinfo {author} {\bibfnamefont {D.}~\bibnamefont {Alonso}},\ } {\bibfield  {journal} {\bibinfo
  {journal} {Phys. Rev. A}\ }\textbf {\bibinfo {volume} {86}},\ \bibinfo
  {pages} {012110} (\bibinfo {year} {2012})}\BibitemShut {NoStop}%
\bibitem{paris2004bose}
M. M. Cola, M. G. A. Paris, and N. Piovella,
Phys. Rev. A {\bf 70}, 043809 (2004).
\bibitem [{\citenamefont {Ferraro}\ and\ \citenamefont
  {Paris}(2005)}]{ferraro20052}%
  \BibitemOpen
  \bibfield  {author} {\bibinfo {author} {\bibfnamefont {A.}~\bibnamefont
  {Ferraro}}\ and\ \bibinfo {author} {\bibfnamefont {M.~G.~A.}\ \bibnamefont
  {Paris}},\ } {\bibfield  {journal}
  {\bibinfo  {journal} {Phys. Rev. A}\ }\textbf {\bibinfo {volume} {72}},\
  \bibinfo {pages} {032312} (\bibinfo {year} {2005})}\BibitemShut {NoStop}%
\bibitem [{\citenamefont {Adesso}\ \emph {et~al.}(2006)\citenamefont {Adesso},
  \citenamefont {Serafini},\ and\ \citenamefont {Illuminati}}]{adesso20061}%
  \BibitemOpen
  \bibfield  {author} {\bibinfo {author} {\bibfnamefont {G.}~\bibnamefont
  {Adesso}}, \bibinfo {author} {\bibfnamefont {A.}~\bibnamefont {Serafini}}, \
  and\ \bibinfo {author} {\bibfnamefont {F.}~\bibnamefont {Illuminati}},\
  } {\bibfield  {journal} {\bibinfo
  {journal} {Phys. Rev. A}\ }\textbf {\bibinfo {volume} {73}},\ \bibinfo
  {pages} {032345} (\bibinfo {year} {2006})}\BibitemShut {NoStop}%
\bibitem [{\citenamefont {Xiang}\ \emph {et~al.}(2009)\citenamefont {Xiang},
  \citenamefont {Shao}, \citenamefont {Song},\ and\ \citenamefont
  {Zou}}]{xiang20091}%
  \BibitemOpen
  \bibfield  {author} {\bibinfo {author} {\bibfnamefont {S.}~\bibnamefont
  {Xiang}}, \bibinfo {author} {\bibfnamefont {B.}~\bibnamefont {Shao}},
  \bibinfo {author} {\bibfnamefont {K.}~\bibnamefont {Song}}, \ and\ \bibinfo
  {author} {\bibfnamefont {J.}~\bibnamefont {Zou}},\ }\href@noop {} {\bibfield
  {journal} {\bibinfo  {journal} {Phys. Rev. A}\ }\textbf {\bibinfo
  {volume} {79}},\ \bibinfo {pages} {032333} (\bibinfo {year}
  {2009})}\BibitemShut {NoStop}%
\bibitem [{\citenamefont {xiang Li}\ \emph {et~al.}(2010)\citenamefont {xiang
  Li}, \citenamefont {hui Sun},\ and\ \citenamefont {Ficek}}]{gao_xiang20101}%
  \BibitemOpen
  \bibfield  {author} {\bibinfo {author} {\bibfnamefont {G.}~\bibnamefont
  {xiang Li}}, \bibinfo {author} {\bibfnamefont {L.}~\bibnamefont {hui Sun}}, \
  and\ \bibinfo {author} {\bibfnamefont {Z.}~\bibnamefont {Ficek}},\ } {\bibfield  {journal}
  {\bibinfo  {journal} {J. Phys. B}\ }\textbf {\bibinfo {volume} {43}},\ \bibinfo {pages} {135501}
  (\bibinfo {year} {2010})}\BibitemShut {NoStop}%
\bibitem [{\citenamefont {Li}\ \emph {et~al.}(2011)\citenamefont {Li},
  \citenamefont {Fogarty}, \citenamefont {Cormick}, \citenamefont {Goold},
  \citenamefont {Busch},\ and\ \citenamefont {Paternostro}}]{li20111}%
  \BibitemOpen
  \bibfield  {author} {\bibinfo {author} {\bibfnamefont {J.}~\bibnamefont
  {Li}}, \bibinfo {author} {\bibfnamefont {T.}~\bibnamefont {Fogarty}},
  \bibinfo {author} {\bibfnamefont {C.}~\bibnamefont {Cormick}}, \bibinfo
  {author} {\bibfnamefont {J.}~\bibnamefont {Goold}}, \bibinfo {author}
  {\bibfnamefont {T.}~\bibnamefont {Busch}}, \ and\ \bibinfo {author}
  {\bibfnamefont {M.}~\bibnamefont {Paternostro}},\ } {\bibfield  {journal} {\bibinfo  {journal} {Phys.
  Rev. A}\ }\textbf {\bibinfo {volume} {84}},\ \bibinfo {pages} {022321}
  (\bibinfo {year} {2011})}\BibitemShut {NoStop}%
\bibitem{galve2013tripartite}
G. Manzano, F. Galve, and R. Zambrini, 
Phys. Rev. A {\bf 87}, 032114 (2013).  
\bibitem [{\citenamefont {Weiss}(1999)}]{weiss1999}%
  \BibitemOpen
  \bibfield  {author} {\bibinfo {author} {\bibfnamefont {U.}~\bibnamefont
  {Weiss}},\ }\href@noop {} {\emph {\bibinfo {title} {Quantum dissipative
  systems}}},\ edited by\ \bibinfo {editor} {\bibfnamefont {W.}~\bibnamefont
  {Scientific}},\ Series in Modern Condensed Matter Physics\ (\bibinfo
  {publisher} {World Scientific},\ \bibinfo {year} {1999})\BibitemShut
  {NoStop}%
\bibitem [{\citenamefont {Brown}\ \emph {et~al.}(2011)\citenamefont {Brown},
  \citenamefont {Ospelkaus}, \citenamefont {Colombe}, \citenamefont {Wilson},
  \citenamefont {Leibfried},\ and\ \citenamefont {Wineland}}]{brown20111}%
  \BibitemOpen
  \bibfield  {author} {\bibinfo {author} {\bibfnamefont {K.}~\bibnamefont
  {Brown}}, \bibinfo {author} {\bibfnamefont {C.}~\bibnamefont {Ospelkaus}},
  \bibinfo {author} {\bibfnamefont {Y.}~\bibnamefont {Colombe}}, \bibinfo
  {author} {\bibfnamefont {A.}~\bibnamefont {Wilson}}, \bibinfo {author}
  {\bibfnamefont {D.}~\bibnamefont {Leibfried}}, \ and\ \bibinfo {author}
  {\bibfnamefont {D.}~\bibnamefont {Wineland}},\ }\href@noop {} {\bibfield
  {journal} {\bibinfo  {journal} {Nature}\ }\textbf {\bibinfo {volume} {471}},\
  \bibinfo {pages} {196} (\bibinfo {year} {2011})}\BibitemShut {NoStop}%
\bibitem [{\citenamefont {Buks}\ and\ \citenamefont
  {Roukes}(2002)}]{buks20021}%
  \BibitemOpen
  \bibfield  {author} {\bibinfo {author} {\bibfnamefont {E.}~\bibnamefont
  {Buks}}\ and\ \bibinfo {author} {\bibfnamefont {M.}~\bibnamefont {Roukes}},\
  }\href@noop {} {\bibfield  {journal} {\bibinfo  {journal}
  {J. Microelectromech. S.}\ }\textbf {\bibinfo {volume}
  {11}},\ \bibinfo {pages} {802} (\bibinfo {year} {2002})}\BibitemShut
  {NoStop}%
\bibitem [{\citenamefont {Cleland}\ and\ \citenamefont
  {Roukes}(2002)}]{cleland20021}%
  \BibitemOpen
  \bibfield  {author} {\bibinfo {author} {\bibfnamefont {A.}~\bibnamefont
  {Cleland}}\ and\ \bibinfo {author} {\bibfnamefont {M.}~\bibnamefont
  {Roukes}},\ }\href@noop {} {\bibfield  {journal} {\bibinfo  {journal}
  {J. Appl. Phys.}\ }\textbf {\bibinfo {volume} {92}},\ \bibinfo
  {pages} {2758} (\bibinfo {year} {2002})}\BibitemShut {NoStop}%
\bibitem [{\citenamefont {Caldeira}\ and\ \citenamefont
  {Leggett}(1983)}]{caldeira1983}%
  \BibitemOpen
  \bibfield  {author} {\bibinfo {author} {\bibfnamefont {A.~O.}\ \bibnamefont
  {Caldeira}}\ and\ \bibinfo {author} {\bibfnamefont {A.~J.}\ \bibnamefont
  {Leggett}},\ }\href@noop {} {\bibfield  {journal} {\bibinfo  {journal}
  {Ann. Phys.}\ }\textbf {\bibinfo {volume} {149}},\ \bibinfo {pages}
  {374} (\bibinfo {year} {1983})}\BibitemShut {NoStop}%
\bibitem [{\citenamefont {Hanggi}\ and\ \citenamefont
  {Ingold}(2005)}]{hanggi20051}%
  \BibitemOpen
  \bibfield  {author} {\bibinfo {author} {\bibfnamefont {P.}~\bibnamefont
  {Hanggi}}\ and\ \bibinfo {author} {\bibfnamefont {G.-L.}\ \bibnamefont
  {Ingold}},\ }\href@noop {} {\bibfield  {journal} {\bibinfo  {journal}
  {Chaos}\ }\textbf {\bibinfo {volume} {15}},\ \bibinfo {pages} {026105}
  (\bibinfo {year} {2005})}\BibitemShut {NoStop}%
\bibitem{breuer}
H. P. Breuer and F. Petruccione, 
\textit{The Theory of Open Quantum Systems}  (Oxford University Press, USA 2012).
\bibitem{garbert1984oscillator}
H. Garbert, U. Weiss, and P. Talkner,
Z. Phys. B \textbf{55}, 87 (1984).
\bibitem{riseborough1895exact}
P. Riseborough, P. Hanggi, and U. Weiss,
Phys. Rev. A \textbf{31}, 471 (1985).
\bibitem{bennet1996eof}
C. H. Bennett, D. P. DiVincenzo, J. A. Smolin, and W. K. Wootters,
Phys. Rev. A \textbf{54}, 3824–3851 (1996).
\bibitem [{\citenamefont {Vidal}\ and\ \citenamefont
  {Werner}(2002)}]{vidal2002a}%
  \BibitemOpen
  \bibfield  {author} {\bibinfo {author} {\bibfnamefont {G.}~\bibnamefont
  {Vidal}}\ and\ \bibinfo {author} {\bibfnamefont {R.~F.}\ \bibnamefont
  {Werner}},\ } {\bibfield
  {journal} {\bibinfo  {journal} {Phys. Rev. A}\ }\textbf {\bibinfo {volume}
  {65}},\ \bibinfo {pages} {032314} (\bibinfo {year} {2002})}\BibitemShut
  {NoStop}%
\bibitem{coffman2000CKW}
V. Coffman, J. Kundu, and W. K. Wootters,
Phys. Rev. A \textbf{61}, 052306 (2000).
\bibitem{peres1996ppt}
A. Peres,
Phys. Rev. Lett. \textbf{77}, 1413 (1996).
\bibitem{horodecki1996ppt}
M. Horodecki, P. Horodecki, and R. Horodecki,
Phys. Lett. A \textbf{223}, 1 (1996).
\bibitem{werner20001gaussian1n}
R. F. Werner and M. M. Wolf,
Phys. Rev. Lett. \textbf{86}, 3658 (2001).
\bibitem [{\citenamefont {Haake}\ and\ \citenamefont
  {Reibold}(1985)}]{haake19851}%
  \BibitemOpen
  \bibfield  {author} {\bibinfo {author} {\bibfnamefont {F.}~\bibnamefont
  {Haake}}\ and\ \bibinfo {author} {\bibfnamefont {R.}~\bibnamefont
  {Reibold}},\ } {\bibfield  {journal}
  {\bibinfo  {journal} {Phys. Rev. A}\ }\textbf {\bibinfo {volume} {32}},\
  \bibinfo {pages} {2462} (\bibinfo {year} {1985})}\BibitemShut {NoStop}%
\bibitem [{\citenamefont {Anders}(2008)}]{anders20081}%
  \BibitemOpen
  \bibfield  {author} {\bibinfo {author} {\bibfnamefont {J.}~\bibnamefont
  {Anders}},\ } {\bibfield
  {journal} {\bibinfo  {journal} {Phys. Rev. A}\ }\textbf {\bibinfo {volume}
  {77}},\ \bibinfo {pages} {062102} (\bibinfo {year} {2008})}\BibitemShut
  {NoStop}%
\bibitem{plenio2008networks}
M. B. Plenio and S. F. Huelga,
New. J. Phys. \textbf{10}, 113019 (2008).
\bibitem [{\citenamefont {Wu}\ and\ \citenamefont {Segal}(2011)}]{wu20111}%
  \BibitemOpen
  \bibfield  {author} {\bibinfo {author} {\bibfnamefont {L.-A.}\ \bibnamefont
  {Wu}}\ and\ \bibinfo {author} {\bibfnamefont {D.}~\bibnamefont {Segal}},\
  } {\bibfield  {journal} {\bibinfo
  {journal} {Phys. Rev. A}\ }\textbf {\bibinfo {volume} {84}},\ \bibinfo
  {pages} {012319} (\bibinfo {year} {2011})}\BibitemShut {NoStop}%
\bibitem{correa2013performance}
L. A. Correa, J. P. Palao, G. Adesso, and D. Alonso,
Phys. Rev. E \textbf{87}, 042131 (2013).
\bibitem{dhar2006transport}
A. Dhar and D. Roy,
J. Stat. Phys. \textbf{127}, 801 (2006).
\bibitem{abrahmowitz1972handbook} 
M. Abramowitz and I. A. Stegun,
\textit{Handbook of Mathematical Functions with Formulas, Graphs, and Mathematical Tables}  (Dover, New York 1972). 
\end{thebibliography}

\end{document}